\documentclass[12pt]{article}

\usepackage{amssymb}
\usepackage{amsmath}
\usepackage{amscd}
\usepackage{graphicx}

\newcommand{\eg}{\textit{e.g.}}
\newcommand{\ie}{\textit{i.e.}}
\newcommand{\sn}{\mathrm{sn}}
\newcommand{\cn}{\mathrm{cn}}
\newcommand{\dn}{\mathrm{dn}}
\newcommand{\SUtwo}{\mathrm{SU}(2)}
\newcommand{\SUN}{\mathrm{SU}(N)}
\newcommand{\Ione}{I_1}
\newcommand{\Itwo}{I_2}

\newcommand{\Pcop}{\mathrm{P_{cop}}}
\newcommand{\Pcol}{\mathrm{P_{col}}}

\def\Vec#1{\mbox{\boldmath $#1$}}

\numberwithin{equation}{section}

\begin{document}
\thispagestyle{empty}
\large
\begin{center}
\textbf{Multi-Calorons Revisited}
\end{center}

\medskip
\begin{center}
Atsushi Nakamula\footnote{e-mail: nakamula@sci.kitasato-u.ac.jp} and Jun Sakaguchi\footnote{e-mail: jsakaguchi@sci.kitasato-u.ac.jp}\\
\bigskip
\emph{Department of Physics, School of Science\\ Kitasato University,\\ Sagamihara, 228-8555, Japan}\\
\end{center}

\normalsize

\vspace{2cm}

\begin{center}
\textbf{Abstract}
\end{center}

\begin{quote}
Analytic Nahm data is re-examined for $\SUtwo$ calorons, or periodic instantons, of instanton charge $2$.
The Nahm equations are solved analytically in terms of Jacobi elliptic functions and the  possible matching conditions are classified.
The dimensions of framed moduli space for charge $2$ caloron is enumerated and
the maximal parameter, which is $16$ for charge $2$, case is identified.
The  monopole and instanton limits are also considered.
It is found that the Nahm data which does not correspond to the standard monopoles plays significant role for calorons.
\end{quote}

%\begin{keyword}
%Calorons \sep ADHM \sep $q$-deformation
%\PACS 02.30.Gp \sep 11.15.-q \sep 11.27.+d
%\end{keyword}
%\end{frontmatter}
\setcounter{page}{0}
\newpage
% main text
\section{Introduction}

%lead
Calorons are finite-action ASD (anti-self-dual) solutions of Yang-Mills gauge theories on partially compactified space $\mathbb{R}^3\times S^1$.
We can interpret them as not only periodic instantons around the $S^1$ direction, but also the composite objects of constituent ``monopoles".
The latter interpretation is due to the fact that calorons can be seen as the monopoles of loop group gauge theories \cite{GM,Norb}.   
Accordingly, they have  the charges of both instanton and monopole.

On analytic solutions of calorons, there has been found  solutions with trivial holonomy in late 1970's \cite{HS} by  ansatz.
A special class of periodic instantons are also found in \cite{Chak}.
In 1980's,  Nahm formulated a systematic device to construct calorons, as well as monopoles, 
by applying the ADHM construction of instantons \cite{ADHM}, \ie, the Nahm construction \cite{Nahm}. 
In this formulation,  caloron gauge connections can be obtained by solving one-dimensional Weyl equations on finite intervals 
with ``impurities" at the boundaries.
The dual gauge connection in Weyl operators, along with the impurities,
 are called Nahm data, which gives the moduli space of calorons.
The dual transformation between the Nahm data and the caloron gauge connections is called the Nahm transform,
 which plays crucial role in the Nahm construction.
The bijection between the Nahm data and the $\SUN$ calorons has been considered earlier in \cite{nye},
and proved in \cite{CH}.

For a long time, calorons have been attracted much attention from various field of mathematical physics.
In gauge theories, particularly important topic on calorons is that they are considered to be explaining 
the confinement phase in finite temperature QCD.
Although calorons are thought to be ineffective for such non-perturbative effects in early days \cite{GPJ},
 the  discovery of  the calorons with non-trivial holonomy \cite{LL, KvB}, at the end of the last century, changes situation drastically. 
They are currently believed to play a fundamental role for the confinement.
In this context, the characteristic properties of calorons, \eg, action densities, quantum corrections \cite{DiaGro, GroSli, Sli} and so on,
 are vigorously studied for $\SUN$ gauge group.
 
Another interest on calorons comes from the string-theoretical point of view, where a dual transformation of D-branes
leads to the Nahm construction of calorons, \ie, the D-brane interpretation \cite{Dia, LY, KapSe}.
Recent development of this field provides further interest on the instanton moduli space on ALF (asymptotically locally flat)
 hyper-K\"ahler four-manifolds \cite{Witten09, Cherkis09}, such as multi-centered Taub-NUT spaces.
Since the trivial, or flat, ALF space is $\mathbb{R}^3\times S^1$,
 exhaustive analysis  of caloron moduli space will give further insight into the instanton moduli space on non-flat ALF spaces.

%nahm data so far
On the  analysis of  the caloron moduli space, it has been shown  that the Nahm data of $\SUN$ calorons are an open subset of 
that of $\SUN$  instantons \cite{nogradi},
 and  that the dimensions of the framed moduli space to $\SUtwo$ calorons are $8k$  \cite{EJ}, where $k$ is an instanton charge.
Although the formal structure of the caloron moduli space is examined in this way,
 the explicit descriptions of the Nahm data, particularly for higher charge cases,
 are studied in quite restricted cases due to analytical difficulty.    

In this paper, we investigate the explicit parameterizations to the Nahm data for $\SUtwo$ calorons of instanton charge $k=2$ in detail,
though the ``general" Nahm data have been appeared already in \cite{nogradi, BNvB03, BNvB04}.
We make a re-construction of the Nahm data in their appropriate form to overview  the caloron moduli space $\mathcal{M}_2$, 
and find that they include the Nahm data of $16$ parameters, which is maximal, \ie, $\mathrm{dim}\mathcal{M}_2$.

%interpolation between R^4 and R^3
We also consider the interrelation between calorons and monopoles or instantons, respectively.
It is naively believed that the large instanton scale limit of calorons gives BPS monopoles in $\mathbb{R}^3$, 
while the large period limit gives instantons in $\mathbb{R}^4$.
In this context,
 it has been shown that there actually exist those limiting cases in the $k=2$ calorons with special symmetry \cite{harland}. 
However, it is known that even for the calorons of $k=2$ with trivial holonomy do not have a monopole limit \cite{ward1}.
Thus, a natural question arises that which conditions on calorons admit the monopole limit.
In this paper, we also give the examples of $k=2$ calorons  which have monopole or instanton limit.

This paper is organized as follows.
In section 2, we make an overview of the Nahm transform for calorons.
In section 3, we give the exact Nahm data of $\SUtwo$ two-calorons in their appropriate form to count the moduli space dimensions.
In section 4, we give the moduli space dimensions for the two-calorons obtained in section 3, which include the maximal parameter cases.
In section 5, we consider the large scale and the large period limit of the calorons obtained so far.
Section 6 is devoted to concluding remarks.

%contents

\section{Nahm transform for calorons}

\subsection{Nahm data of calorons}
%the topological charges
One of the characteristic charge of calorons is the instanton charge $k$.
As mentioned in introduction, calorons can be interpreted as a composite object of  ``constituent monopoles".
In $\SUN$ gauge theory, the constituent monopoles are composed of $N$-set of multi-monopoles.
In this paper, we restrict ourselves to consider calorons of $\SUtwo$ gauge group without net magnetic charge.
Of this type of calorons, the two-set of multi-monopoles have an equal and opposite magnetic charges.
We occasionally denote the charge $k$-calorons without net magnetic charge as charge $(k,k)$-calorons.

Another significant quantity for calorons is the masses of constituent monopoles, $(\nu_1,\nu_2)=(2\mu,\mu_0-2\mu)$,
 where $\mu$ and $\mu_0$ are defined shortly. 
Note that the order of the constituent monopoles is not crucial, because there is a large gauge transformation which exchange this order,
the rotation map.
The asymptotic form of the gauge potential for calorons is characterized by these  caloron charges and the net magnetic charge $k_M$, as
\begin{equation}
A_0=i\mathrm{diag}(\mu,-\mu)-\frac{i}{2r}\mathrm{diag}(k_M,-k_M)+O\left(r^{-2}\right).
\end{equation}
We consider only the $k_M=0$ case as mentioned above.

The caloron Nahm data consists of two parts.
One of them is the bulk Nahm data, and the other is the boundary data.
The bulk Nahm data is four hermitian $k\times k$ matrices $T_\alpha(s)\,(\alpha=0,1,2,3)$, which are
 analytic and periodic in $s$ with periodicity $\mu_0$, and defined on the fundamental interval $I=(-\mu,\mu_0-\mu)$.
The fundamental interval is composed of two parts, $I=I_1\oplus I_2=(-\mu,\mu)\oplus (\mu, \mu_0-\mu)$. 
We denote the four matrices $T_\alpha(s)=T^{(1)}_\alpha(s)$ on $I_1$ and $T_\alpha(s)=T^{(2)}_\alpha(s)$  on $I_2$, respectively.  
Note that the definition of the fundamental interval and its decomposition are equivalent to $I=(-\mu_0/2,\mu_0/2)$, and
$I_1=(-\mu,\mu)$ and $I_2=(-\frac{\mu_0}{2},-\mu)\oplus(\mu, \frac{\mu_0}{2})$, by the periodicity.

For the gauge field to be anti-self-dual, the matrices must  enjoy the Nahm equations on each interval,
\begin{equation}
\frac{d}{ds}T_a^{(m)}-i\left[T^{(m)}_0\,,T^{(m)}_a\right]-\frac{i}{2}\epsilon_{abc}\left[T^{(m)}_b\,,T^{(m)}_c\right]=0,\label{caloronNahm}
\end{equation}
where $m=1$ or $2$, and the roman subscripts are $1,\,2$ or $3$.
They also have to satisfy the reality conditions
\begin{equation}
T^{(m)}_{\alpha}(-s)={}^tT^{(m)}_{\alpha}(s),\label{reality}
\end{equation}
where the periodicity is taken into account.

The boundary Nahm data are a $k$-row vector $W$ with quaternion element, and quaternion projection matrices
\begin{equation}
P_\pm=\frac{1}{2}\left(1\pm i\Vec{\omega}\cdot\Vec{e}\right),
\end{equation}
where $\Vec{\omega}\cdot\Vec{e}$ is a pure imaginary unit quaternion, namely, $|\Vec{\omega}|=1$.
Here the definition of  the quaternion element $e_\alpha$ is
\begin{equation}
(e_0,e_1,e_2,e_3):=(1,\Vec{e})=(1,-i\sigma_1,-i\sigma_2,-i\sigma_3).
\end{equation}
They give the matching conditions for the bulk Nahm data at $s=\pm\mu$, such as
\begin{subequations}\label{matching}
\begin{eqnarray}
T_j^{(2)}(\mu)-T_j^{(1)}(\mu)=\frac{1}{2}\mbox{Tr}_2\left(\sigma_j W^\dagger P_+ W\right)\label{matchingat+}\\
T_j^{(1)}(-\mu)-T_j^{(2)}(\mu_0-\mu)=\frac{1}{2}\mbox{Tr}_2\left(\sigma_j W^\dagger P_- W\right),\label{matchingat-}
\end{eqnarray}
\end{subequations}
for the $(k,k)$-calorons, where $j=1,2$ or $3$.
In the right-hand-side of (\ref{matching}), the trace is for quaternions so that they give a $k\times k$ matrix.

For later purpose, we show the exact form of the right-hand-side of (\ref{matching}) for $k=2$,
 which can be expressed  in terms of the standard $u(2)$  basis.
Let us take $W=\left(p, q\right)$  with two quaternions $p$ and $q$, whose components are $p=p_0+\Vec{p}\cdot\Vec{e}$
 and $q=q_0+\Vec{q}\cdot\Vec{e}$.
Then we find
\begin{eqnarray}
&\frac{1}{2}\mbox{Tr}_2\left(\sigma_j W^\dagger P_\pm W\right)\nonumber\\
&=\pm S_j\sigma_1+A_j\sigma_2 \pm \frac{1}{2}\left(M_1-M_2\right)_j\sigma_3
\pm\frac{1}{2}\left(M_1+M_2\right)_j\,1_2,\label{matchingk=2}
\end{eqnarray}
where the coefficients of $u(2)$ basis are
\begin{subequations}\label{generalmatching}
\begin{eqnarray}
&S_j=&\frac{1}{2}\Bigl(p_0q_0\,\omega_j+p_0\left(\Vec{\omega}\times\Vec{q}\right)_j+q_0\left(\Vec{\omega}\times\Vec{p}\right)_j \nonumber\\
&{}&+\left(\Vec{\omega}\cdot\Vec{p}\right)\,q_j-\left(\Vec{p}\cdot\Vec{q}\right)\,\omega_j+\left(\Vec{\omega}\cdot\Vec{q}\right)\,p_j\Bigr)\label{Sj}\\
&A_j=&\frac{1}{2}\Bigl(p_0q_j-q_0p_j-\left(\Vec{p}\times\Vec{q}\right)_j\Bigr)\label{Aj}\\
&M_{1,j}=&\frac{1}{2}\left(p_0^2-\Vec{p}^2\right)\omega_j+p_0\left(\Vec{\omega}\times\Vec{p}\right)_j
+\left(\Vec{\omega}\cdot\Vec{p}\right)\,p_j \label{M1j}\\
&M_{2,j}=&\frac{1}{2}\left(q_0^2-\Vec{q}^2\right)\omega_j+q_0\left(\Vec{\omega}\times\Vec{q}\right)_j
+\left(\Vec{\omega}\cdot\Vec{q}\right)\,q_j.\label{M2j}
\end{eqnarray}
\end{subequations}

\subsection{Nahm transform}
As is well known,  a caloron  gauge field is obtained from the Nahm data through the Nahm transform.
The procedure of the Nahm transform for the $\SUtwo$ calorons is as follows.
Firstly, we solve the ``Weyl equation" for $k$-column ``spinor" $U^{(m)}(s;x^\alpha)$ with quaternion entries defined on each interval $I_1$ and $I_2$,
\begin{equation}
\left\{\frac{d}{ds}-i\left(T^{(m)}_0(s)+T^{(m)}_j(s)e_j+x\right)\right\}U^{(m)}(s;x^\alpha)=0,
\end{equation}
where  $x=x_0+\Vec{x}\cdot\Vec{e}$.
Next, we find out a quaternion $V(x^\alpha)$ enjoying the matching conditions
\begin{subequations}
\begin{eqnarray}
U^{(2)}(\mu; x^\alpha)-U^{(1)}(\mu; x^\alpha)=iW^\dagger P_+V(x^\alpha)\\
U^{(1)}(-\mu;x^\alpha)-U^{(2)}(\mu_0-\mu;x^\alpha)=iW^\dagger P_-V(x^\alpha).
\end{eqnarray}
\end{subequations}
In addition,  $U(s;x^\alpha)$ and $V(x^\alpha)$ should be normalized as 
\begin{equation}
\int_{I_1} \left.U^{(1)}\right.^\dagger U^{(1)} ds+\int_{I_2}\left.U^{(2)}\right.^\dagger U^{(2)} ds+V^\dagger V=1_2.
\end{equation}
By using these ``parts", the gauge potential is given by
\begin{equation}
A_\alpha=\int_{I_1}\left.U^{(1)}\right.^\dagger \partial_\alpha U^{(1)} ds
+\int_{I_2}\left.U^{(2)}\right.^\dagger \partial_\alpha U^{(2)} ds+V^\dagger \partial_\alpha V.\label{gaugepotential}
\end{equation}

\subsection{Gauge transformation to Nahm data}

There exist continuous transformations for the Nahm data which preserve the gauge potential (\ref{gaugepotential}).
One of them is generated by a $U(2)$ valued smooth function  $g(s)$ defined on $I$,
enjoying reality condition $g^\dagger(s)=g(-s)$.
The $g(s)$ action on the Nahm data is 
\begin{subequations}\label{gaugetrf}
\begin{eqnarray}
&T^{(m)}_j(s) \mapsto g(s)T^{(m)}_j(s) g^{-1}(s)\label{gtforT_j}\\
&T^{(m)}_0(s) \mapsto g(s)T^{(m)}_0(s) g^{-1}(s)+ig(s)\dfrac{d}{ds}g^{-1}(s)\label{gaugetrfT0}\\
&W\mapsto P_+Wg^{-1}(\mu)+P_-Wg^{-1}(-\mu).
\end{eqnarray}
\end{subequations}
Simultaneously, the Weyl spinor transforms as
\begin{equation}
U^{(m)}(s)\mapsto g(s)U^{(m)}(s)
\end{equation}
with $V$ unchanged.

There is another transformation generated by a unit quaternion $h$ which preserves (\ref{gaugepotential}).
It acts on the Nahm data as
\begin{subequations}
\begin{eqnarray}
W \mapsto hW\\
\Vec{\omega}\cdot\Vec{e} \mapsto h\,\Vec{\omega}\cdot\Vec{e}\, h^\dagger,
\end{eqnarray}
\end{subequations}
and
\begin{equation}
V\mapsto hV.
\end{equation}

We call these two types of transformation as the gauge transformation for the Nahm data.

\section{Exact Nahm data  of two-calorons}
In this section, we will find the analytic Nahm data of $k=2$ caloron in its appropriate form for later consideration.

\subsection{Bulk solutions}
First, we have to solve the bulk equations (\ref{caloronNahm})  in both intervals $\Ione$ and $\Itwo$, independently.
We take the following form for the $2\times2$ matrices $T^{(m)}_{\alpha}(s)$ by using the standard $u(2)$ basis,
\begin{subequations}\label{Nahmansatz}
\begin{eqnarray}
&T^{(m)}_a(s)=&f^{(m)}_1(s)\sigma_1\quad +g^{(m)}_1(s)\sigma_3+d^{(m)}_1(s)1_2\\
&T^{(m)}_b(s)=&\qquad\ f^{(m)}_2(s)\sigma_2 \qquad \qquad +d^{(m)}_2(s)1_2\label{sigma_2}\\
&T^{(m)}_c(s)=&g^{(m)}_3(s)\sigma_1\quad +f^{(m)}_3(s)\sigma_3+d^{(m)}_3(s)1_2,
\end{eqnarray}
\end{subequations}
where $(a,b,c)$ is a cyclic permutation of $(1,2,3)$, $f^{(m)}_j(s)$, $g^{(m)}_j(s)$ and $d^{(m)}_j(s)$ are functions on each interval.
In addition, we assume $T^{(m)}_0=d^{(m)}_01_2$, where $d^{(m)}_0$ is a constant to be a moduli parameter. 
From (\ref{Nahmansatz}), the Nahm equations (\ref{caloronNahm}) are
\begin{subequations}\label{diffeqns}
\begin{eqnarray}
\dot{f}^{(m)}_1=-2f^{(m)}_2 f^{(m)}_3, \qquad \qquad \dot{g}^{(m)}_1=2f^{(m)}_2g^{(m)}_3\\
 \dot{f}^{(m)}_2=-2\left(f^{(m)}_1f^{(m)}_3-g^{(m)}_1g^{(m)}_3\right)\\
\dot{g}^{(m)}_3=2f^{(m)}_2 g^{(m)}_1, \qquad \qquad \dot{f}^{(m)}_3=-2f^{(m)}_2f^{(m)}_1,
\end{eqnarray}
\end{subequations}
and $\dot{d}^{(m)}_j=0 \ (j=1,2,3)$, which lead to constant $d^{(m)}_j$'s.
Note that we do not specify the ordering of the indices $(a,b,c)$ in the left-hand-side of (\ref{Nahmansatz}) at this stage,
because any cyclic permutations of $(1,2,3)$ gives the same system of differential equations (\ref{diffeqns}).
Clearly, this ordering ambiguity does not affect the monopole Nahm data.
However, the relative difference of the ordering between each interval gives physical significance for caloron Nahm data, as we will see later.

The system (\ref{diffeqns}) can be solved by elliptic functions.
Since (\ref{diffeqns}) is invariant under the exchange of $g^{(m)}_1$ and $g^{(m)}_3$ (and also $f^{(m)}_1$ and $f^{(m)}_3$), 
 we find that there are two distinct type of solutions, which can not be transformed each other by a ``gauge transformation".  
The first type of the solution is in the following form 
\begin{subequations}\label{monopole}
\begin{eqnarray}
&f^{(m)}_1(s)=a^{(m)}(s)\cos\phi_m \quad g^{(m)}_1(s)=-a^{(m)}(s)\sin\phi_m,\label{monopole-a}\\
&g^{(m)}_3(s)=\pm b^{(m)}(s)\sin\phi_m \quad  f^{(m)}_3(s)=\pm b^{(m)}(s)\cos\phi_m,\label{monopole-b}\\
&f^{(m)}_2(s)=\mp D_m k_m'\dfrac{\sn2D_m(s-s_m)}{\cn2D_m(s-s_m)}\label{monopole-c},
\end{eqnarray}
\end{subequations}
where the functions $a^{(m)}(s)$ and $b^{(m)}(s)$ are 
\begin{eqnarray}
\left(a^{(m)}(s),b^{(m)}(s)\right)=\left(D_m\frac{k_m'}{\cn2D_m(s-s_m)}, D_m\frac{\dn2D_m(s-s_m)}{\cn2D_m(s-s_m)}\right)\nonumber\\
 \mbox{or}\ \left(D_m\frac{\dn2D_m(s-s_m)}{\cn2D_m(s-s_m)}, D_m\frac{k_m'}{\cn2D_m(s-s_m)}\right).\label{a and b}
\end{eqnarray}
Here $\sn,\,\cn$ and $\dn$ are Jacobi elliptic functions of modulus $k_m$ (and $k_m':=\sqrt{1-k_m^2}$) and $D_m,\; \phi_m,$ and $s_m$ are constants.
The ``origin", or the location of symmetric axis, $s_m$, is chosen to  enjoy the reality conditions (\ref{reality}), namely
 $s_1=0$  and $s_2=\mu_0/2$, respectively.
For  this solution, if $D_m$ is set to the complete elliptic integral $K(k_m)$, with $s\in (-1,1)$ and $\,s_m=0$, then
 we find that the solution satisfies the $\SUtwo$ monopole boundary conditions \cite{Hit}.
In fact, each function has simple poles at the boundaries $s=1$ and $-1$, 
and their residues give an irreducible representation of $su(2)$.
Also, they give rise to the ordinary monopole spectral curve.
Thus we find (\ref{monopole}) corresponds to the Nahm data of two-monopole in this choice of parameters.
For this reason, we refer to the solution (\ref{monopole}) as ``monopole type", abbreviated as ``M".

The other type of solution is 
\begin{subequations}\label{non-m}
\begin{eqnarray}
&f^{(m)}_1(s)=a^{(m)}(s)\cos\phi_m\quad g^{(m)}_1(s)=-b^{(m)}(s)\sin\phi_m,\label{non-m-a}\\
&g^{(m)}_3(s)=\pm a^{(m)}(s)\sin\phi_m \quad f^{(m)}_3(s)=\pm b^{(m)}(s)\cos\phi_m,\label{non-m-b}\\
&f^{(m)}_2(s)=\mp D_mk_m'\dfrac{\sn2D_m(s-s_m)}{\cn2D_m(s-s_m)},\label{non-m-c}
\end{eqnarray}
\end{subequations}
where the definitions of $\left(a^{(m)}(s),b^{(m)}(s)\right)$ are (\ref{a and b}).
For this type, we easily find that they do not give a standard monopole Nahm data, but a spatially rotated case, in the situation discussed above.
Accordingly, they do not give a monopole spectral curve simply.
Thus, we refer to this solution as ``non-standard monopole" type, abbreviated as ``non-M".
Although this type of solutions are irrelevant for the monopole Nahm data, we will find that they are considerable for the caloron Nahm data.

The bulk Nahm data are given by a combination of the solutions in each interval $\Ione$ and $\Itwo$:
there are three types of combinations to the bulk data, (non-M, non-M), (non-M, M) and (M, M),
where the order in the combination of the solutions is not crucial due to the rotation map.

\subsection{Classification of matching conditions}
As we mentioned in the previous section, the boundary Nahm data, which gives the matching conditions of the bulk data,
are the  two-component row vector $W$ with quaternion entry and the projection matrices $P_\pm$.
In this subsection, we give their exact form and find that there are two distinct types of matching conditions,
 ``parallel" and ``orthogonal", for the $k=2$ calorons.

Firstly, we fix the first component of the row vector $W$ being real by using the gauge transformation, \ie,
\begin{equation}
W=\left(\lambda, \rho\hat{q}\right),
\end{equation}
where $\lambda,\;\rho\in \mathbb{R}$ and $\hat{q}:=\hat{q}_\mu e_\mu$ is a unit quaternion.
Here we take the parameterization of $\hat{q}$ by using  $S^3$ coordinate as
\begin{eqnarray}
&\hat{q}_0=\cos\psi\label{q_0},\\
&{}^t\hat{\Vec{q}}:=
\left[\begin{array}{c}
\hat{q}_1\\
\hat{q}_2\\
\hat{q}_3\\
\end{array}\right]
=\left[
\begin{array}{l}
\sin\psi \sin \theta\sin\varphi\\
\sin\psi \sin \theta\cos\varphi\\
\sin\psi \cos \theta\\
\end{array}\right]\label{qhat}
\end{eqnarray}
where $0\leq\psi,\,\theta\leq\pi$ and $0\leq\varphi\leq2\pi$.
In this parameterization, the vectors in (\ref{generalmatching}) reduce to
\begin{subequations}\label{reducedmatching}
\begin{eqnarray}
&S_j=&\frac{1}{2}\lambda\rho\left(\omega_j+\left(\Vec{\omega}\times\hat{\Vec{q}}\right)_j\right)\label{reducedSj}\\
&A_j=&\frac{1}{2}\lambda\rho\, \hat{q}_j\label{reducedAj}\\
&M_{1,j}=&\frac{1}{2}\lambda^2 \omega_j\label{reducedD1j}\\
&M_{2,j}=&\frac{1}{2}\rho^2\cos2\psi \,\omega_j+\rho^2\cos\psi\left(\Vec{\omega}\times\hat{\Vec{q}}\right)_j
+\rho^2\Vec{\omega}\cdot\hat{\Vec{q}}\,\hat{q}_j\label{reducedD2j}
\end{eqnarray}
\end{subequations}
Next, we take the parameterization of $\Vec{\omega}$ in the projection matrices $P_\pm$,  according to
the following two types of matching conditions.

\subsubsection{Parallel type matching}
The first  matching type is ``parallel type", in which the bulk Nahm data $T_j^{(1)}$ and $T_j^{(2)}$ have the same ``$\sigma_2$-direction":
the $\sigma_2$ directions (\ref{sigma_2}) of  both $T_j^{(1)}$ and  $T_j^{(2)}$ agree, or are parallel.
In this case, we can describe the matching conditions (\ref{matchingat+}) in terms of the following $3\times3$ matrix,   
\begin{eqnarray}
\left[
\begin{array}{ccc}
f^{(2)}_1(\mu) & 0 & g^{(2)}_1(\mu)\\
0 & f^{(2)}_2(\mu) & 0\\
g^{(2)}_3(\mu)& 0 & f^{(2)}_3(\mu)\\
\end{array}
\right]
-\left[
\begin{array}{ccc}
f^{(1)}_1(\mu) & 0 & g^{(1)}_1(\mu)\\
0 & f^{(1)}_2(\mu) & 0\\
g^{(1)}_3(\mu) & 0 & f^{(1)}_3(\mu)\\
\end{array}
\right]\nonumber\\
=\left[
\begin{array}{ccc}
S_1 & A_1 & U_1\\
S_2 & A_2 & U_2\\
S_3 & A_3 & U_3\\
\end{array}
\right],\label{matchingmatrixP}
\end{eqnarray}
where the column index is specified by the direction indexed by $j$, and the row index is specified by the direction of $su(2)$ basis.
Here we have defined  
\begin{equation}
U_j:=\dfrac{1}{2}(M_{1,j}-M_{2,j}),\ (j=1,2,3).\label{U_j}
\end{equation}
Note that the matching conditions at $s=-\mu$, (\ref{matchingat-}), are automatically satisfied due to the reality condition $T_j(-s)={}^tT_j(s)$.
From (\ref{matchingmatrixP}), we find that the components $A_1, A_3, S_2$ and $U_2$ have to be zero. 
Thus, we can separate (\ref{matchingmatrixP}) into two parts: one of them is  
\begin{equation}
f^{(2)}_2(\mu)-f^{(1)}_2(\mu)=A_2,\label{f_2matching}
\end{equation}
and the other part is given by $2\times2$ matrix, 
\begin{eqnarray}
&\left[
\begin{array}{cc}
f^{(2)}_1(\mu)  & g^{(2)}_1(\mu)\\
g^{(2)}_3(\mu)& f^{(2)}_3(\mu)\\
\end{array}
\right]
-\left[
\begin{array}{cc}
f^{(1)}_1(\mu)  & g^{(1)}_1(\mu)\\
g^{(1)}_3(\mu)  & f^{(1)}_3(\mu)\\
\end{array}
\right]=\left[
\begin{array}{cc}
S_1  & U_1\\
S_3  & U_3\\
\end{array}
\right].\label{2by2}
\end{eqnarray}

From (\ref{qhat}) and (\ref{reducedAj}), the conditions $A_1=A_3=0$ lead to 
\begin{equation}
\hat{\Vec{q}}=(0,\sin\psi,0),
\end{equation}
which is equivalent to $\theta=\frac{\pi}{2}$ and $\varphi=0$ or $\pi$.
Thus, we take the parameterization, 
\begin{equation}
\Vec{\omega}=\left(\sin\xi \sin\eta,\,\cos\xi,\,\sin\xi \cos\eta\right),
\end{equation}
which leads to
\begin{equation}
\Vec{\omega}\times\hat{\Vec{q}}=\sin\psi\,\left(-\sin\xi\cos\eta,0,\sin\xi\sin\eta\right).
\end{equation} 

Next, we can easily find that the conditions $S_2=U_2=0$ are given by two distinct ways.
Firstly, from (\ref{reducedSj}), (\ref{reducedD1j}), (\ref{reducedD2j}) and (\ref{U_j}), we find it is sufficient to choose 
\begin{equation}
\xi=\frac{\pi}{2}. \label{coplanar}
\end{equation}
In this case, the coefficients of the unit matrix part in the bulk data $d^{(m)}_j$ are subject to the constraints
\begin{eqnarray}
&{}^t\Delta\Vec{d}:=\left[
\begin{array}{c}
d^{(2)}_1-d^{(1)}_1\\
d^{(2)}_2-d^{(1)}_2\\
d^{(2)}_3-d^{(1)}_3
\end{array}
\right]\nonumber\\
&=\left[
\begin{array}{c}
\frac{1}{4}\left\{(\lambda^2+\rho^2\cos2\psi)\sin\eta-\rho^2\sin2\psi\cos\eta\right\}\\
0\\
\frac{1}{4}\left\{(\lambda^2+\rho^2\cos2\psi)\cos\eta+\rho^2\sin2\psi\sin\eta\right\}
\end{array}
\right].
\label{distance-coplanar}
\end{eqnarray}
Thus, from the analogy of monopole Nahm data,
 we find that the spatial separation of the ``centers" of each two-monopoles lies on $(1,3)$-plane,
 and that the ``$2$-axes" of each two-monopoles are located in parallel and coplanar in some plane perpendicular to $(1,3)$-plane.
We refer this type of matching condition as ``parallel and coplanar", abbreviated as $\mathrm{P_{cop}}$.
The non-zero components of the right-hand-side of (\ref{matchingmatrixP}) are 
\begin{subequations}\label{Pcop-components}
\begin{eqnarray}
&A_2=\pm\frac{1}{2}\lambda\rho\sin\psi\label{Pcop-A2}\\
&S_1=\frac{1}{2}\lambda\rho\sin(\eta-\psi)\label{Pcop-S1}\\
&S_3=\frac{1}{2}\lambda\rho\cos(\eta-\psi)\label{Pcop-S3}\\
&U_1=\frac{1}{4}\left(\lambda^2\sin\eta-\rho^2\sin(\eta-2\psi)\right)\label{Pcop-U1}\\
&U_3=\frac{1}{4}\left(\lambda^2\cos\eta-\rho^2\cos(\eta-2\psi)\right),\label{Pcop-U3}
\end{eqnarray}
\end{subequations} 
where the sign of $A_2$ is chosen appropriately according to the choice of the bulk solutions.
Note that a part of this matching conditions corresponds to the ``crossed" configuration considered in \cite{BNvB04}.

The other choice to realize the constraints  $S_2=U_2=0$ is 
\begin{equation}
\lambda=\rho, \ \mbox{and}\quad \psi=\frac{\pi}{2}.\label{collinear}
\end{equation}
In this case, the constraints on $d^{(m)}_j$ are
\begin{eqnarray}
\Delta\Vec{d}:=\left(
d^{(2)}_1-d^{(1)}_1,
d^{(2)}_2-d^{(1)}_2,
d^{(2)}_3-d^{(1)}_3\right)
=\left(0,\frac{1}{2}\lambda^2\cos\xi,0\right).\label{distance-collinear}
\end{eqnarray}
We find that the centers of the two-monopoles are both on the ``$2$-axis",
 which means that the main axes of both monopoles are collinear.
Thus, we refer this type of matching condition as ``parallel and collinear", abbreviated as $\mathrm{P_{col}}$.
The non-zero components of the right-hand-side of (\ref{matchingmatrixP}) are 
\begin{subequations}\label{Pcol-components}
\begin{eqnarray}
&A_2=\pm\frac{1}{2}\lambda^2\label{Pcol-A2}\\
&S_1=-\frac{1}{2}\lambda^2\sin\xi\cos\eta\label{Pcol-S1}\\
&S_3=\frac{1}{2}\lambda^2\sin\xi\sin\eta\label{Pcol-S3}\\
&U_1=\frac{1}{2}\lambda^2\sin\xi\sin\eta\label{Pcol-U1}\\
&U_3=\frac{1}{2}\lambda^2\sin\xi\cos\eta\label{Pcol-U3}.
\end{eqnarray}
\end{subequations} 
A part of this matching condition is studied in \cite{harland},
 and corresponds to the ``rectangular" configuration considered in \cite{BNvB04}.

\subsubsection{Orthogonal type matching}

The other matching type is ``orthogonal type", in which the $\sigma_2$-directions of $T_j^{(1)}$ and  $T_j^{(2)}$ are orthogonal. 
Without loss of generality, we can take the $\sigma_2$-direction of  $T_j^{(1)}$ is $j=2$  and  that of $T_j^{(2)}$ is $j=1$, respectively.
In this case, the matching conditions (\ref{matchingat+}) are given by the following $3\times3$ matrix, 
\begin{eqnarray}
\left[
\begin{array}{ccc}
0 & f^{(2)}_2(\mu) & 0\\
g^{(2)}_3(\mu)& 0 & f^{(2)}_3(\mu)\\
f^{(2)}_1(\mu) & 0 & g^{(2)}_1(\mu)\\
\end{array}
\right]
-\left[
\begin{array}{ccc}
f^{(1)}_1(\mu) & 0 & g^{(1)}_1(\mu)\\
0 & f^{(1)}_2(\mu) & 0\\
g^{(1)}_3(\mu) & 0 & f^{(1)}_3(\mu)\\
\end{array}
\right]\nonumber\\
=\left[
\begin{array}{ccc}
S_1 & A_1 & U_1\\
S_2 & A_2 & U_2\\
S_3 & A_3 & U_3\\
\end{array}
\right],\label{matchingmatrixO}
\end{eqnarray}
where $U_j$ is defined by (\ref{U_j}) as in the previous case.
For this type of matching conditions,
 the permanently zero component of the right-hand-side is only $A_3$, which leads to $\theta=\pi/2$ from (\ref{qhat}).
Hence, we find
\begin{equation}
\hat{\Vec{q}}=(\sin\psi\sin\varphi, \sin\psi\cos\varphi,0).
\end{equation}
The components in the right-hand-side of matching conditions (\ref{matchingmatrixO}) are summarized to
\begin{subequations}\label{O-components}
\begin{eqnarray}
&A_1=\frac{1}{2}\lambda\rho\sin\psi\sin\varphi\\
&A_2=\frac{1}{2}\lambda\rho\sin\psi\cos\varphi\\
&\Vec{S}:=(S_1,S_2,S_3)=
\frac{1}{2}\lambda\rho\left(\cos\psi\,\Vec{\omega}-\sin\psi\,\Vec{\omega}\times\hat{\Vec{q}}\right)\label{Sdef}\\
&\Vec{U}:=(U_1,U_2,U_3)
=\frac{1}{4}\left(\lambda^2-\rho^2\cos2\psi\right)\Vec{\omega}+\frac{1}{4}\rho^2\sin2\psi\,\Vec{\omega}\times\hat{\Vec{q}} \nonumber\\
&-\frac{1}{2}\rho^2\sin^2\psi\ \left(\Vec{\omega}\cdot\hat{\Vec{q}}\right)\hat{\Vec{q}},\label{Udef}
\end{eqnarray}
\end{subequations} 
where, we have fixed the parameterization of $\Vec{\omega}$ as
\begin{equation}
\Vec{\omega}=(\sin\xi\cos\eta, \sin\xi\sin\eta, \cos\xi).
\end{equation}
for later convenience.
In this type of matching conditions, the separation of the center of two-monopoles is 
\begin{eqnarray}
&\Delta\Vec{d}:=\left(d_1^{(2)}-d_1^{(1)},d_2^{(2)}-d_2^{(1)},d_3^{(2)}-d_3^{(1)}\right)\nonumber\\
&=\frac{1}{4}\left(\lambda^2+\rho^2\cos2\psi\right)\Vec{\omega}-\frac{1}{4}\rho^2\sin2\psi\,\Vec{\omega}\times\hat{\Vec{q}} 
+\frac{1}{2}\rho^2\sin^2\psi\ \left(\Vec{\omega}\cdot\hat{\Vec{q}}\right)\hat{\Vec{q}}.\label{distance-O}
\end{eqnarray}
In contrast to the type-P matching, any component of $\Delta\Vec{d}$ is  generally non-zero.

To conclude this section,  we have made a classification of $\SUtwo$ calorons of instanton charge $k=2$ by the two criterion:
the first is the type of bulk Nahm data and the second is the type of boundary data, the matching conditions.
The bulk Nahm data are classified into three distinct cases, (non-M, non-M), (non-M, M) and (M, M).
Of the boundary data, the possible matching types are the pararell type (type-P) and the orthogonal type (type-O),
 and the former is further classified into the pararell and coplanar type (type-$\mathrm{P_{cop}}$) and the 
pararell and collinear type (type-$\mathrm{P_{col}}$).
In the next section,
 we consider the moduli space dimensions of $k=2$ calorons by investigating all of the combination of the bulk and boundary data
discussed above.

\section{Moduli space dimensions}

In this section, we implement definite counting of the moduli space dimensions to the caloron Nahm data obtained so far.
As mentioned in introduction, it is proved that the framed moduli space dimensions of  $\SUtwo$ calorons of charge $k$ are $8k$ \cite{EJ}.
We will find that the (non-M, non-M) bulk solutions with the type-$\Pcop$ matching condition 
  give the case of maximal dimensions for $k=2$, \ie, the 16-dimensional framed moduli space.

\subsection{The (non-M, non-M) bulk data}

First of all, we consider the bulk solutions of the type-(non-M, non-M) associated to respective matching conditions.
We firstly investigate the type-P matching conditions (\ref{f_2matching}) and  (\ref{2by2}), and enumerate the moduli space dimensions for both
type-$\Pcop$ and type-$\Pcol$ matching conditions, followed by the consideration on the type-O matching conditions.

\subsubsection{Type-$\Pcop$ matching conditions}

From the non-M type solutions (\ref{non-m}),  the matrix part (\ref{2by2}) can be written as
\begin{eqnarray}
&\left[
\begin{array}{cc}
a^{(2)}(\mu)\cos\phi_2  & -b^{(2)}(\mu)\sin \phi_2\\
a^{(2)}(\mu)\sin\phi_2& b^{(2)}(\mu)\cos\phi_2\\
\end{array}
\right]
-\left[
\begin{array}{cc}
a^{(1)}(\mu)\cos\phi_1  & -b^{(1)}(\mu)\sin \phi_1\\
a^{(1)}(\mu)\sin\phi_1& b^{(1)}(\mu)\cos\phi_1\\
\end{array}
\right]\nonumber\\
&=\left[
\begin{array}{cc}
S_1  & U_1\\
S_3  & U_3\\
\end{array}
\right],\label{2by2ab}
\end{eqnarray}
where $a^{(m)}$ and $b^{(m)}$ are defined earlier in (\ref{a and b}).
We can find that (\ref{2by2ab}) is regarded as vector relations in $\mathbb{R}^2$,  by defining
\begin{subequations}\label{a and b in Pcop}
\begin{eqnarray}
\Vec{a}^{(m)}:=a^{(m)}(\mu)
\left(\begin{array}{c}
\cos \phi_m\\
\sin \phi_m\\
\end{array}
\right),\\
\Vec{b}^{(m)}:=b^{(m)}(\mu)
\left(\begin{array}{c}
-\sin \phi_m\\
\cos\phi_m\\
\end{array}
\right)
\end{eqnarray}
\end{subequations}
and, from (\ref{Pcop-S1}) -- (\ref{Pcop-U3}),
\begin{eqnarray}
&\left(\begin{array}{c}
S_1\\
S_3\\
\end{array}
\right)=\dfrac{\lambda\rho}{2}
\left(
\begin{array}{c}
\sin (\eta-\psi)\\
\cos (\eta-\psi)\\
\end{array}
\right)
=:\Vec{k}_{\mathrm{cop}}
,\label{kcop}\\
&\left(\begin{array}{c}
U_1\\
U_3\\
\end{array}
\right)
=:\Vec{l}_{\mathrm{cop}}-\Vec{m}_{\mathrm{cop}},
\end{eqnarray}
where
\begin{equation}
\Vec{l}_{\mathrm{cop}}
=\dfrac{\lambda^2}{4}
\left(
\begin{array}{c}
\sin \eta\\
\cos \eta\\
\end{array}
\right),\
\Vec{m}_{\mathrm{cop}}=\dfrac{\rho^2}{4}
\left(
\begin{array}{c}
\sin (\eta-2\psi)\\
\cos (\eta-2\psi)\\
\end{array}
\right).\label{lmcop}
\end{equation}
Thus, we recast (\ref{2by2ab}) to,
\begin{eqnarray}
&\Vec{a}^{(2)}-\Vec{a}^{(1)}=\Vec{k}_{\mathrm{cop}}\label{a2-a1-Pcop}\\
&\Vec{b}^{(2)}-\Vec{b}^{(1)}=\Vec{l}_{\mathrm{cop}}-\Vec{m}_{\mathrm{cop}}\label{b2-b1-Pcop}.
\end{eqnarray}
Note that each vector is given in the polar representation in $\mathbb{R}^2$ and
 that $\Vec{k}_{\mathrm{cop}}$ bisects the angle between $\Vec{l}_{\mathrm{cop}}$ and $\Vec{m}_{\mathrm{cop}}$.
In addition to these relations, there is another constraint (\ref{f_2matching}), which has a form 
\begin{equation}
-D_2k_2'\dfrac{\;\sn2D_2(\mu-\frac{\mu_0}{2})\;}{\cn2D_2(\mu-\frac{\mu_0}{2})}
+D_1k_1'\dfrac{\sn2D_1\mu}{\;\cn2D_1\mu\;}=\frac{1}{2}\lambda\rho\sin\psi,\label{PcopA2matching}
\end{equation}
by (\ref{non-m-c}).
Here we have selected the bulk solutions on both intervals with the minus sign case of  (\ref{non-m-c}).
Although it is possible to consider the other combination of the sign in (\ref{non-m-c}), the following analysis can be achieved similarly.
Hence, we fix the matching condition as (\ref{PcopA2matching}).
By graphical analysis, we find the relations (\ref{a2-a1-Pcop}) and (\ref{b2-b1-Pcop}) have solutions with continuous parameters.
When $\phi_2-\phi_1=0$, both $\Vec{a}^{(2)}$ and $\Vec{a}^{(1)}$,
 and $\Vec{b}^{(2)}$ and $\Vec{b}^{(1)}$ have the same directions so that the vector relations are shown as in Figure 1.
When $\phi_2-\phi_1\neq0$, those are shown as in Figure 2. 

\begin{figure}
\includegraphics[width=10cm]{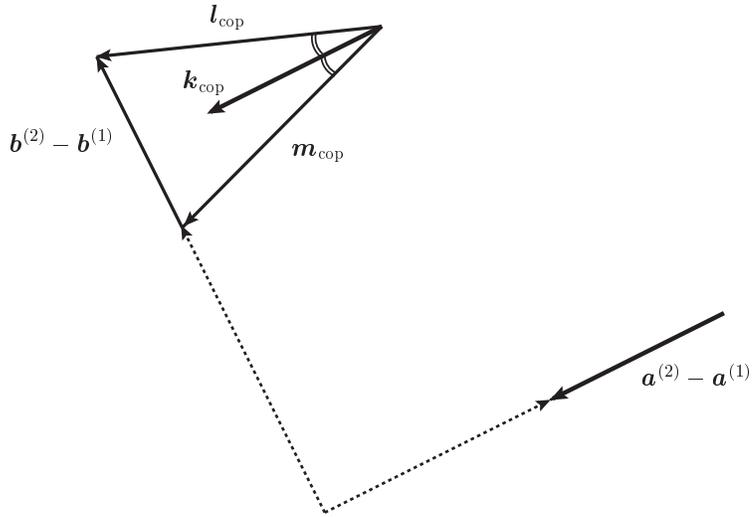}
\caption{The case $\phi_2-\phi_1=0$.}
\label{fig1}
\end{figure}

\begin{figure}
\includegraphics[width=10cm]{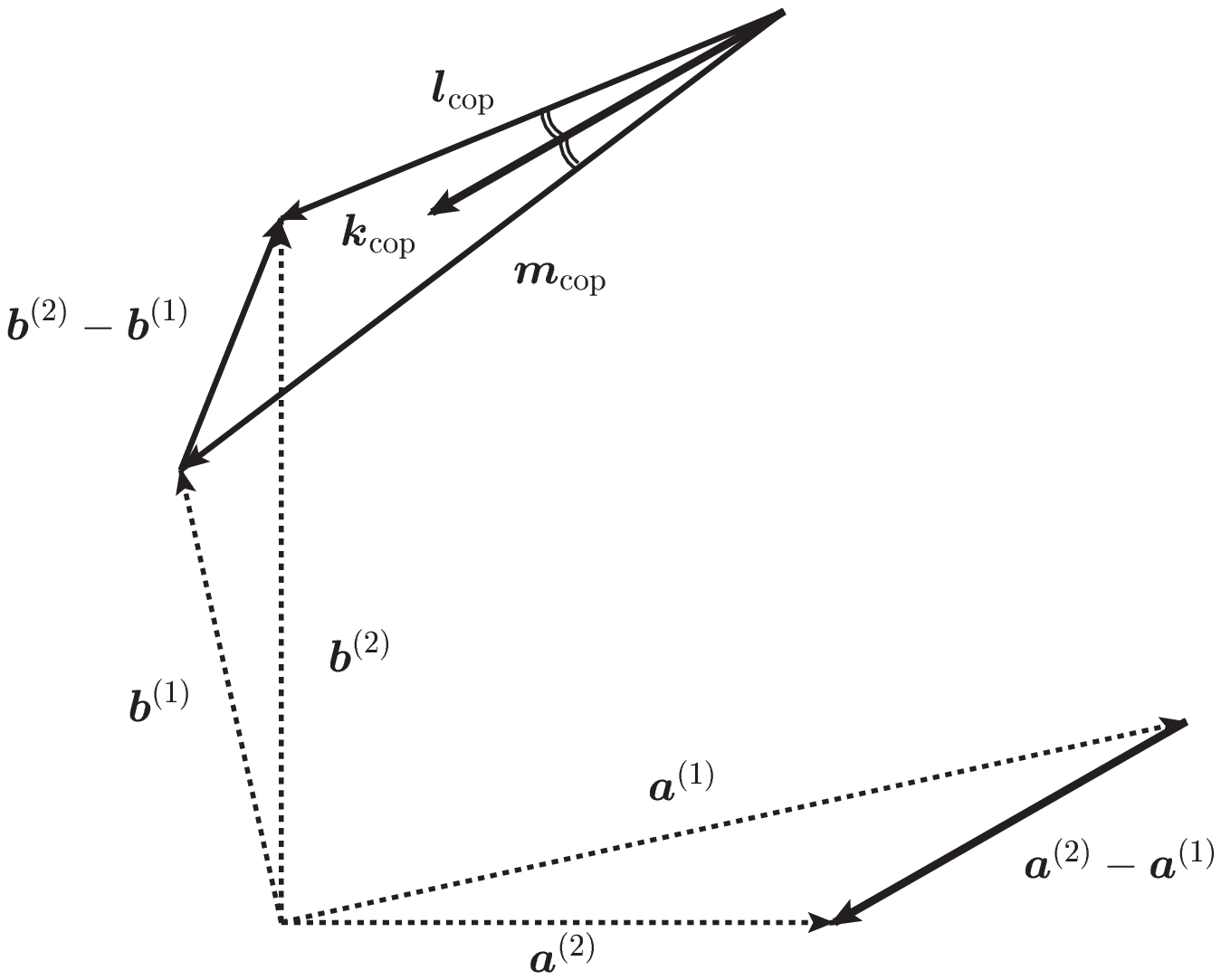}
\caption{The case $\phi_2-\phi_1\neq0$.}
\label{fig2}
\end{figure}

%graphical explanation

The moduli space dimensions, \ie, the number of continuous parameters minus the number of constraints,
 for the type-$\Pcop$ matching can be obtained by explicit counting of them.
In this case, the parameters in the solution are the following 16:
\begin{equation}
\lambda, \rho, \psi, \eta, k_1, k_2, D_1, D_2, \phi_1, \phi_2, d^{(1)}_0, d^{(2)}_0, d^{(1)}_1, d^{(1)}_2, d^{(1)}_3, \mu, \label{Pcopmoduli}
\end{equation}
and the constraints on these parameters are (\ref{a2-a1-Pcop}), (\ref{b2-b1-Pcop}) and (\ref{PcopA2matching}), 
The first two constraints are the constraints on vectors in $\mathbb{R}^2$, so these are respectively two constraints in general.
However, all of the vectors are given by polar representation, in which the origins of the angle $\phi_m$'s and $\eta$ 
can be freely chosen, thus we can always let one component in each constraint $0=0$.
Hence, the constraints (\ref{a2-a1-Pcop}) and (\ref{b2-b1-Pcop}) are the only relations for the norm of both sides, respectively.
In fact, the second components of (\ref{a2-a1-Pcop}) and (\ref{b2-b1-Pcop}) are obtained from the first components of them by
the shift of the angle $\phi_m\to\phi_m-\frac{\pi}{2}$ and $\eta\to\eta+\frac{\pi}{2}$, and vice versa.  
Therefore, we conclude that the number of constraints are 3 for this case: the number of free parameter is $16-3=13$.
In addition to this, there should be a degree of freedom of global gauge rotation, which has 3 independent parameters for framed moduli
space as in the case of instantons in $\mathbb{R}^4$.
Therefore, the moduli space dimensions for $\Pcop$ is $13+3=16$,  which is the maximal number to $k=2$ calorons.

\subsubsection{Type-$\Pcol$ matching conditions}

Next, we consider the type-$\Pcol$ matching conditions, whose boundary  data are given by (\ref{Pcol-components}). 
In this case, the matrix part of the matching conditions (\ref{2by2}) can be recast to 
\begin{eqnarray}
&\Vec{a}^{(2)}-\Vec{a}^{(1)}=\Vec{k}_{\mathrm{col}}\label{a2-a1-Pcol},\\
&\Vec{b}^{(2)}-\Vec{b}^{(1)}=\Vec{l}_{\mathrm{col}}\label{b2-b1-Pcol},
\end{eqnarray}
where the vectors in the right-hand-side are, from (\ref{Pcol-S1}) -- (\ref{Pcol-U3}),
\begin{eqnarray}
&\Vec{k}_{\mathrm{col}}=\dfrac{\lambda^2\sin\xi}{2}
\left(
\begin{array}{c}
-\cos \eta\\
\sin \eta\\
\end{array}
\right),\label{kforPcol}\\
&\Vec{l}_{\mathrm{col}}=\dfrac{\lambda^2\sin\xi}{2}
\left(
\begin{array}{c}
\sin \eta\\
\cos \eta\\
\end{array}
\right).\label{lforPcol}
\end{eqnarray}
In addition, from (\ref{non-m-c}) and (\ref{Pcol-A2}), we find
\begin{equation}
-D_2k_2'\dfrac{\;\sn2D_2(\mu-\frac{\mu_0}{2})\;}{\cn2D_2(\mu-\frac{\mu_0}{2})}
+D_1k_1'\dfrac{\sn2D_1\mu}{\;\cn2D_1\mu\;}=\frac{1}{2}\lambda^2.\label{PcolA2matching}
\end{equation}

We now show that $\sin\xi=0$ is necessary for the type-$\Pcol$ matching conditions, \ie, $\xi$ is not a moduli parameter. 
If we assume that $\sin\xi\neq0$, then it is found from (\ref{kforPcol}) and (\ref{lforPcol}) that $|\Vec{k}|=|\Vec{l}|$ 
and $\Vec{l}$ is orthogonal to $\Vec{k}$ by clockwise $\pi/2$ rotation.
However, as can be seen in Figure 3, this is not allowed configuration due to the unfavorable orientation.
%\begin{center}
\begin{figure}
\includegraphics[width=10cm]{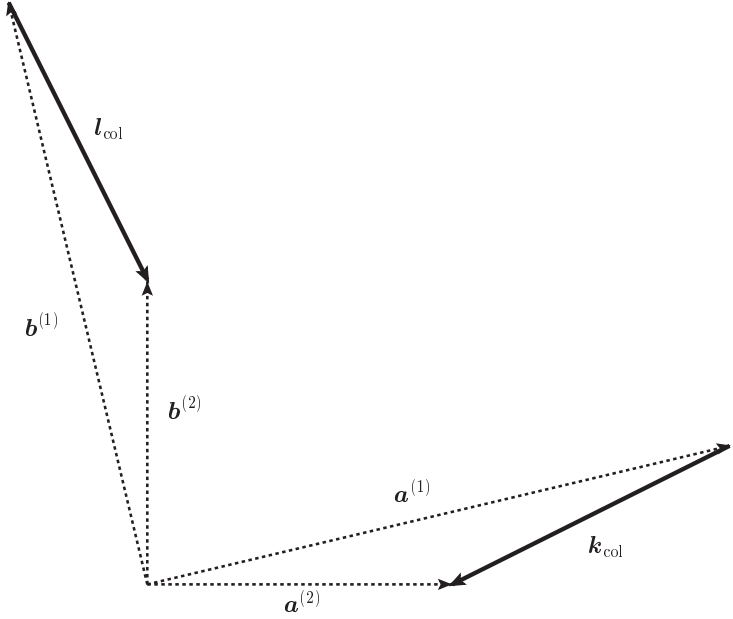}
\caption{The $\sin\xi\neq0$ case.}
\label{fig3}
\end{figure}
%\end{center}

Hence,  it is acceptable only $\sin\xi=0$ for the type-$\Pcol$ matching.
In this case, the matching conditions (\ref{a2-a1-Pcol}) and (\ref{b2-b1-Pcol}) reduce to $\Vec{a}_1=\Vec{a}_2$ and $\Vec{b}_1=\Vec{b}_2$.
They are fulfilled by $\phi_1=\phi_2=\phi$ and,
\begin{subequations}\label{Pcolab-1}
\begin{eqnarray}
\frac{D_1k_1'}{\cn2D_1\mu}=\frac{D_2k_2'}{\cn2D_2(\mu-\frac{\mu_0}{2})},\label{a1=a2}\\
\frac{D_1\cn2D_1\mu}{\dn2D_1\mu}=\frac{D_2\cn2D_2(\mu-\frac{\mu_0}{2})}{\dn2D_2(\mu-\frac{\mu_0}{2})},\label{b1=b2}
\end{eqnarray}
\end{subequations}
or
\begin{subequations}\label{Pcolab-2}
\begin{eqnarray}
\frac{D_1k_1'}{\cn2D_1\mu}=\frac{D_2\cn2D_2(\mu-\frac{\mu_0}{2})}{\dn2D_2(\mu-\frac{\mu_0}{2})}\\
\frac{D_1\cn2D_1\mu}{\dn2D_1\mu}=\frac{D_2k_2'}{\cn2D_2(\mu-\frac{\mu_0}{2})}.
\end{eqnarray}
\end{subequations}
Having eliminated redundant parameters, we find that the  moduli parameters for this configuration are 
\begin{equation}
\lambda, D_1, D_2, k_1, k_2, \phi, d^{(1)}_0, d^{(2)}_0, d^{(1)}_1, d^{(1)}_2, d^{(1)}_3, \mu, \label{Pcolgaplessmoduli}
\end{equation}
with constraints (\ref{Pcolab-1}) or (\ref{Pcolab-2}), and (\ref{PcolA2matching}).
Thus, the framed moduli space dimensions of the type-$\Pcol$ matching is $12-3+3=12$. 
Note that the special case $k_1=k_2=0$ of the type-$\Pcol$ matching conditions  is investigated in detail in \cite{harland}.

\subsubsection{Type-O matching conditions}
The third type of matching conditions is the orthogonal type (\ref{matchingmatrixO}).
For this case, we find the matching conditions  can be regarded as vector relations in $\mathbb{R}^3$,
along with  two independent relations on $f^{(m)}_2$'s, in contrast to the type-P matching . 
That is, if we define
\begin{eqnarray}
\Vec{a}^{(2)}=a^{(2)}(\mu)
\left(
\begin{array}{c}
0\\
\sin \phi_2\\
\cos \phi_2\\
\end{array}
\right),\
\Vec{a}^{(1)}=a^{(1)}(\mu)
\left(\begin{array}{c}
\cos \phi_1\\
0\\
\sin \phi_1\\
\end{array}
\right),\\
\Vec{b}^{(2)}=b^{(2)}(\mu)
\left(\begin{array}{c}
0\\
\cos \phi_2\\
-\sin \phi_2\\
\end{array}
\right),\
\Vec{b}^{(1)}=b^{(1)}(\mu)
\left(\begin{array}{c}
-\sin \phi_1\\
0\\
\cos \phi_1\\
\end{array}
\right),
\end{eqnarray}
then the first and the third columns of (\ref{matchingmatrixO}) are equivalent to
\begin{eqnarray}
&\Vec{a}^{(2)}-\Vec{a}^{(1)}={}^t\Vec{S}\label{a2-a1-O}\\
&\Vec{b}^{(2)}-\Vec{b}^{(1)}={}^t\Vec{U}\label{b2-b1-O},
\end{eqnarray}
where the three-vectors $\Vec{S}$ and $\Vec{U}$ are defined by (\ref{Sdef}) and (\ref{Udef}), respectively.
In (\ref{a2-a1-O}) and (\ref{b2-b1-O}), all the three components are independent each other, because they are not in the special form 
as in the type-P matching conditions.
The rest of the matching conditions are,
\begin{eqnarray}
&\mp D_2k_2'\dfrac{\;\sn2D_2(\mu-\frac{\mu_0}{2})\;}{\cn2D_2(\mu-\frac{\mu_0}{2})}=\frac{1}{2}\lambda\rho\sin\psi\sin\varphi\label{O-A1matching}\\
&\pm D_1k_1'\dfrac{\sn2D_1\mu}{\;\cn2D_1\mu\;}=\frac{1}{2}\lambda\rho\sin\psi\cos\varphi.\label{O-A2matching}
\end{eqnarray}
Hence, we have 8 independent constraints.

We can observe that there actually exist solutions to this type of matching conditions.
For example, if we take $\phi_1=\phi_2=0,  \psi=\eta, \varphi=\eta=\pi/4$,
 then the matching conditions for $\Vec{a}^{(m)}$ and $\Vec{b}^{(m)}$ turn out to be
\begin{subequations}\label{type-O example}
\begin{eqnarray}
a^{(2)}(\mu)=\frac{\rho^2}{\sqrt{2}}\sin\psi\cos^2\psi\\
b^{(2)}(\mu)=\frac{\rho^2}{\sqrt{2}}\cos^2\psi\\
a^{(1)}(\mu)=-\frac{\rho^2}{2}\cos\psi\\
b^{(1)}(\mu)=-\rho^2\sin\psi\cos\psi,
\end{eqnarray}
\end{subequations}
together with $\lambda=\sqrt{2}\rho$.
It can be found that there exists a parameter region for $D_m, k_m\,(i=1,2)$ and $\mu$ which enjoys (\ref{type-O example}) provided that $\pi/2<\psi<\pi$.

In general, for the type-O matching conditions, the parameters in the solutions are the following 18,
\begin{equation}
\lambda, \rho, \psi, \varphi, \xi, \eta, k_1, k_2, D_1, D_2, \phi_1, \phi_2, d^{(1)}_0, d^{(2)}_0, d^{(1)}_1, d^{(1)}_2, d^{(1)}_3, \mu, \label{Omoduli}
\end{equation}
subject to the 8 constraints (\ref{O-A1matching}), (\ref{O-A2matching}), (\ref{a2-a1-O}) and (\ref{b2-b1-O}).
Therefore, the dimension of the framed moduli space for the type-O matching conditions are $18-8+3=13$, where the $+3$ comes from 
a global gauge transformation.

\subsection{The (non-M, M) and (M, M) bulk data}

In this subsection, we consider the cases that at least one of the bulk solution is the monopole type,
namely, the (non-M, M) and (M, M) bulk solutions.
Note that, for the (M, M) bulk data, we can always perform a ``diagonalization" on one of the interval $\Ione$, say,
 by using a gauge transformation (\ref{gtforT_j}).
That is, we can fix the $su(2)$ part of the bulk data on the interval $\Ione$ to be
\begin{subequations}\label{diagmonopole}
\begin{eqnarray}
&T^{(1)}_1(s)=a^{(1)}(s)\;\sigma_1,\label{diagmonopole-1}\\
&T^{(1)}_2(s)=\mp D_1k_1'\dfrac{\sn2D_1s}{\cn2D_1s}\;\sigma_2\label{diagmonopole-2},\\
&T^{(1)}_3(s)=\pm b^{(1)}(s)\;\sigma_3,\label{diagmonopole-3}
\end{eqnarray}
\end{subequations}
for the (M, M) solutions.

\subsubsection{Type-$\Pcop$ matching conditions}
The type-P matching conditions of the matrix part (\ref{2by2}) for the (non-M, M) bulk data are, by using (\ref{non-m}), (\ref{monopole}),
\begin{eqnarray}
\left[
\begin{array}{cc}
a^{(2)}(\mu)\cos\phi_2  & -a^{(2)}(\mu)\sin \phi_2\\
b^{(2)}(\mu)\sin\phi_2& b^{(2)}(\mu)\cos\phi_2\\
\end{array}
\right]
&-\left[
\begin{array}{cc}
a^{(1)}(\mu)\cos\phi_1  & -b^{(1)}(\mu)\sin \phi_1\\
a^{(1)}(\mu)\sin\phi_1& b^{(1)}(\mu)\cos\phi_1\\
\end{array}
\right]\nonumber\\
&=\left[
\begin{array}{cc}
S_1  & U_1\\
S_3  & U_3\\
\end{array}
\right],\label{nonM-M-2by2ab}
\end{eqnarray}
while for the (M, M) bulk data are
\begin{eqnarray}
\left[
\begin{array}{cc}
a^{(2)}(\mu)\cos\phi_2  & -a^{(2)}(\mu)\sin \phi_2\\
b^{(2)}(\mu)\sin\phi_2& b^{(2)}(\mu)\cos\phi_2\\
\end{array}
\right]
-\left[
\begin{array}{cc}
a^{(1)}(\mu)  & 0\\
0  & b^{(1)}(\mu)\\
\end{array}
\right]\nonumber\\
=\left[
\begin{array}{cc}
S_1  & U_1\\
S_3  & U_3\\
\end{array}
\right],\label{M-M-2by2ab}
\end{eqnarray}
where $\phi_2$ is re-defined after the gauge transformation which diagonalize the bulk data on $\Ione$.
From this consideration, we find that the (M, M) data are obtained by restricting $\phi_1=0$ in the (non-M, M) solutions.
Hence, it is sufficient to consider the (non-M, M) solutions.

As in the previous subsection, we firstly consider the type-$\Pcop$ matching conditions.
By using the 2-vectors (\ref{a and b}), (\ref{kcop}) and (\ref{lmcop}), we can recast (\ref{nonM-M-2by2ab}) to, 
\begin{eqnarray}
\tilde{\Vec{a}}^{(2)}-\Vec{a}^{(1)}=\Vec{k}_{\mathrm{cop}},\ 
\label{non-M-M-Pcop1}\\
\tilde{\Vec{b}}^{(2)}-\Vec{b}^{(1)}=\Vec{l}_{\mathrm{cop}}-\Vec{m}_{\mathrm{cop}},\label{non-M-M-Pcop2}
\end{eqnarray}
where
\begin{equation}
\tilde{\Vec{a}}^{(2)}=\left(
\begin{array}{c}
a^{(2)}(\mu)\cos \phi_2\\
b^{(2)}(\mu)\sin \phi_2\\
\end{array}
\right),\
\tilde{\Vec{b}}^{(2)}=\left(
\begin{array}{c}
-a^{(2)}(\mu)\sin \phi_2\\
b^{(2)}(\mu)\cos \phi_2\\
\end{array}
\right).
\end{equation}
Here, the ``vectors" $\tilde{\Vec{a}}^{(2)}$ and $\tilde{\Vec{b}}^{(2)}$ are not in polar representations of vectors in $\mathbb{R}^2$, 
in contrast to the (non-M, non-M) solutions.
As a consequence, there are 2 independent constraints in both (\ref{non-M-M-Pcop1}) and (\ref{non-M-M-Pcop2}). 
We find that the total number of constraints in the type-$\Pcop$ matching is 2+2+1=5, where 1 comes  from (\ref{PcopA2matching}), and that
the parameters in the solution are the same as (\ref{Pcopmoduli}), except for $\phi_1$ for the (M, M) solution.
Hence, the moduli space dimensions to this case are 16-5+3=14 for the (non-M, M) solution and 13 for (M, M) solution.

\subsubsection{Type-$\Pcol$ matching conditions}
Next, we consider the type-$\Pcol$ matching conditions,
which are obtained by replacing (\ref{Pcop-components}) with (\ref{Pcol-components}), as for the (non-M, non-M) bulk data.
The matching conditions are (\ref{PcolA2matching}) along with
\begin{eqnarray}
\tilde{\Vec{a}}^{(2)}-\Vec{a}^{(1)}=\Vec{k}_{\mathrm{col}},\label{MMPcol-a}\\
\tilde{\Vec{b}}^{(2)}-\Vec{b}^{(1)}=\Vec{l}_{\mathrm{col}},\label{MMPcol-b}
\end{eqnarray}
where the right-hand-sides are given by (\ref{kforPcol}) and (\ref{lforPcol}).
In this case, the graphical situation is similar to Figure 3 so that it is necessary $\sin \xi=0$ as in the (non-M, non-M) case.
Hence, we find (\ref{MMPcol-a}) and (\ref{MMPcol-b}) are equivalent to
\begin{subequations}\label{MMPcol}
\begin{eqnarray}
a^{(2)}(\mu)\cos\phi_2-a^{(1)}(\mu)\cos\phi_1=0,\\
b^{(2)}(\mu)\sin\phi_2-a^{(1)}(\mu)\sin\phi_1=0,\\
-a^{(2)}(\mu)\sin\phi_2+b^{(1)}(\mu)\sin\phi_1=0,\\
b^{(2)}(\mu)\cos\phi_2-b^{(1)}(\mu)\cos\phi_1=0,
\end{eqnarray}
\end{subequations}
for the (non-M, M) case.
These four constraints are independent each other as the type-$\Pcop$ matching conditions, so that there are totally $4+1=5$ constraints.
The parameters in the solution to this matching conditions are the following $13$,
\begin{equation}
\lambda, k_1, k_2, D_1, D_2, \phi_1, \phi_2, d^{(1)}_0, d^{(2)}_0, d^{(1)}_1, d^{(1)}_2, d^{(1)}_3, \mu.\label{nm-m-Pcolmoduli}
\end{equation}
Accordingly, we find the dimensions for framed moduli space are $13-5+3=11$.

As mentioned before, we can fix $\phi_1=0$ for the (M, M) case, which  induces $\phi_2=0$ from (\ref{MMPcol}).
Thus, the moduli parameters are  the following $11$,
\begin{equation}
\lambda, k_1, k_2, D_1, D_2, d^{(1)}_0, d^{(2)}_0, d^{(1)}_1, d^{(1)}_2, d^{(1)}_3, \mu. \label{nm-m-Pcolmoduli}
\end{equation}
The moduli space dimensions are $11-5+3=9$ for  the (M, M) case, consequently.

\subsubsection{Type-O matching conditions}
Finally, we consider the matching conditions of the type-O.
As in subsection 4.1.3, we can recast the matching conditions into the relations in $\mathbb{R}^3$.
From the bulk data (\ref{monopole}) and (\ref{non-m}), and the matching conditions (\ref{matchingmatrixO}), we find 
\begin{subequations}\label{nonM-M-O}
\begin{eqnarray}
\left(
\begin{array}{c}
-a^{(1)}(\mu)\cos \phi_1\\
b^{(2)}(\mu)\sin \phi_2\\
a^{(2)}(\mu)\cos \phi_2-a^{(1)}(\mu)\sin \phi_1\\
\end{array}
\right)=
{}^t\Vec{S},\label{non-M-M-OS}\\
\left(
\begin{array}{c}
b^{(1)}(\mu)\sin \phi_1\\
b^{(2)}(\mu)\cos \phi_2\\
-a^{(2)}(\mu)\sin \phi_2-b^{(1)}(\mu)\cos \phi_1\\
\end{array}
\right)=
{}^t\Vec{U},\label{non-M-M-OU}
\end{eqnarray}
\end{subequations}
for the (non-M, M) solution, where $\Vec{S}$ and $\Vec{U}$ are defined by (\ref{Sdef}) and (\ref{Udef}), respectively.
The matching conditions for the (M, M) solution are obtained by restricting $\phi_1=0$, as in the previous subsection.

The number  of independent constraints for these cases is $8$, where $6$ comes from (\ref{nonM-M-O}),
and  others from (\ref{O-A1matching}) and (\ref{O-A2matching}).  
The number of parameters in the solutions are the same as (\ref{Omoduli}), namely, $18$.
Accordingly, the framed moduli space dimensions are $18-8+3=13$ for the (non-M, M) case, and $12$ for the (M, M) case.

\subsection{Summary of this section}

In this section, we have considered the framed moduli space dimensions for diverse types of matching  conditions, which are summarized 
in the following table:
\begin{center}
\begin{tabular}{|cc|c|c|c|}
\hline
 &  Matching type & \multicolumn{2}{c|}{P} & O\\ 
 \cline{3-4}
Bulk solutions  &  & $\mathrm{P_{cop}}$ & $\mathrm{P_{col}}$ &  \\
\hline
(non-M, non-M) &  & 16& 12 & 13\\
\hline
(non-M, M) & & 14& 11& 13\\
\hline
(M, M) &  & 13 & 9 & 12\\
\hline
\end{tabular}
\end{center}

We have found the $\SUtwo$ calorons with $16$-dimensional moduli space appear when the bulk solutions are of the type (non-M, non-M) 
and the matching conditions are of the type-$\Pcop$.

\section{The monopole and instanton limits}
In this section, we consider  the large scale limit, $\lambda,\,\rho\to\infty$, and the large period limit, $\mu_0\to0$, of the caloron Nahm data
obtained so far.
Naively, they correspond to  monopoles in $\mathbb{R}^3$ and instantons in $\mathbb{R}^4$, respectively.   
Thus calorons are expected to interpolate between monopoles in $\mathbb{R}^3$ and instantons in $\mathbb{R}^4$.
However, as we will see in this section, things are somewhat complicated.

\subsection{The monopole limit}
It can easily  be found that  the separation between the center of two ``2-monopoles" diverges at the large scale limit.
In fact, from (\ref{distance-coplanar}) and (\ref{distance-O}), we find  that $|\Delta\Vec{d}|\to\infty$ in the type-$\Pcop$ and the type-O matching 
at the limit of $\lambda\to\infty$ and/or $\rho\to\infty$.
Also from (\ref{distance-collinear}), $\lambda\to\infty$ leads $|\Delta\Vec{d}|\to\infty$ in the type-$\Pcol$ matching.
Simultaneously, the parameters in the bulk Nahm data have to be
\begin{eqnarray}
2D_1\mu\to K_1,\\
2D_2\left|\mu-\frac{\mu_0}{2}\right|\to K_2,
\end{eqnarray}
according to the matching conditions.

In these large scale limit, there remains only the bulk Nahm data on one of the intervals $T^{(1)}_a\,(a=0,1,2,3)$, say.
Hence, if  $T^{(1)}_a$  is the monopole type solution (\ref{monopole}), then we find the BPS two-monopole configuration in this limit.
On the other hand, if  $T^{(1)}_a$ is the non-standard monopole type solution (\ref{non-m}), 
we obtain the BPS two-monopole limit in its spatial rotated one.
However, there should be another type of the large scale limit, the (2,1)-calorons, in general.
To find  the procedure to induce such kind of the large scale limits is under investigation.

\subsection{The instanton limit}
As mentioned above, it is expected that calorons turn out to be instantons on $\mathbb{R}^4$, in the large period limit $\mu_0\to0$. 
In this subsection, we will show that there are some restrictions on the caloron Nahm data which reproduces instanton limit.
We have two approaches to obtain  the instanton ADHM data from the caloron Nahm data with non-trivial holonomy.
The first approach is that we take $\mu_0\to0$ with the ``monopole mass ratio", $c:=\nu_2/\nu_1$, fixed during the process.
In this approach, it is necessary that the bulk Nahm data of both intervals coincide, as we will see.
The other approach is that we take $\mu_0\to0$ via the calorons with trivial holonomy.
Namely, if we firstly take the limit $\mu\to\frac{\mu_0}{2}$ (or $\mu\to0$ by the rotation map) to let one of the two-monopole be massless,
 then we obtain the Harrington-Shepard type calorons.
It is known that those calorons have instanton limit without restriction.
Hence, we only consider the first approach here.

Firstly, we consider  the calorons of type-P matching conditions at the limit $\mu_0\to0$.
The relations between each monopole mass and the period $\mu_0$ are given in terms of the monopole mass ratio $c$ as,
\begin{equation}
\nu_1=\mu=\frac{\mu_0}{2(1+c)},\ \nu_2=\frac{\mu_0}{2}-\mu=\frac{c\mu_0}{2(1+c)}.
\end{equation}
where $c$ is defined as
\begin{equation}
c:=\frac{\nu_2}{\nu_1}=\frac{\mu_0-2\mu}{2\mu}.
\end{equation}
It is observed that the period $\mu_0$ and the scale parameters of calorons, $\lambda$ and $\rho$ for type-$\Pcop$,  and $\lambda$ for type-$\Pcol$,
 are linked by the matching condition (\ref{f_2matching}). 
By taking the $\mu_0\to0$ limit, the left-hand-side of (\ref{f_2matching}) turns out to be
\begin{eqnarray}
-D_2k_2'\dfrac{\;\sn2D_2(\mu-\frac{\mu_0}{2})\;}{\cn2D_2(\mu-\frac{\mu_0}{2})}
+D_1k_1'\dfrac{\sn2D_1\mu}{\;\cn2D_1\mu\;}\nonumber\\
=D_2k_2'\dfrac{\;\sn2D_2\nu_2\;}{\cn2D_2\nu_2}
+D_1k_1'\dfrac{\sn2D_1\nu_1}{\;\cn2D_1\nu_1\;}\nonumber\\
\xrightarrow[\mu_0\to0]{} \ \frac{\mu_0}{2(1+c)}\left(cD_2^2k_2'+D_1^2k_1'\right),\label{f_2LHSlimit}
\end{eqnarray}
while the right-hand-side is given by (\ref{Pcop-A2}) for the type-$\Pcop$ and (\ref{Pcol-A2}) for the type-$\Pcol$ matching conditions.
This means that the scale parameters $\lambda$ and/or $\rho$ must tend to be zero when $\mu_0\to0$.
Accordingly, we find that all the components in the right-hand-side of (\ref{2by2}) are zero when $\mu_0\to0$,
from (\ref{Pcop-components}) for the type-$\Pcop$ or (\ref{Pcol-components}) for the type-$\Pcol$ matching conditions, respectively.
Thus, we observe that  (\ref{2by2}) with $\phi_1=\phi_2$ turns out to be 
\begin{subequations}\label{coincide}
\begin{eqnarray}
\lim_{\mu_0\to0}\left(\frac{D_2k_2'}{\cn2D_2(\mu-\frac{\mu_0}{2})}-\frac{D_1k_1'}{\cn2D_1\mu}\right)
=D_2k_2'-D_1k_1'=0\\
\lim_{\mu_0\to0}\left(\frac{D_2\dn2D_2(\mu-\frac{\mu_0}{2})}{\cn2D_2(\mu-\frac{\mu_0}{2})}-\frac{D_1\dn2D_1\mu}{\cn2D_1\mu}\right)
=D_2-D_1= 0
\end{eqnarray}
\end{subequations}
or
\begin{subequations}\label{non-coincide}
\begin{eqnarray}
\lim_{\mu_0\to0}\left(\frac{D_2k_2'}{\cn2D_2(\mu-\frac{\mu_0}{2})}-\frac{D_1\dn2D_1\mu}{\cn2D_1\mu}\right)
=D_2k_2'-D_1=0\\
\lim_{\mu_0\to0}\left(\frac{D_2\dn2D_2(\mu-\frac{\mu_0}{2})}{\cn2D_2(\mu-\frac{\mu_0}{2})}-\frac{D_1k_1'}{\cn2D_1\mu}\right)
=D_2-D_1k_1'= 0
\end{eqnarray}
\end{subequations}
according to the bulk Nahm data (\ref{monopole}) or (\ref{non-m}), and (\ref{a and b}).
Note that if $\phi_1\neq\phi_2$ then we find there is no solution for $D_m$ and $k_m$ in this limit. 
Hence,  we conclude  that all the parameters in the bulk Nahm data for both intervals have to coincide, \ie, $\phi_1=\phi_2=:\phi$, 
$D_1=D_2=:D$ and $k_1=k_2=:k$ at $\mu_0\to 0$.
In particular,  $k=0$ is necessary  for the case (\ref{non-coincide}).

Consequently, from (\ref{f_2matching}) and (\ref{f_2LHSlimit}), the $\mu_0$ dependence of the scale parameters is 
\begin{subequations}
\begin{eqnarray}
\lambda\sim\tilde{\lambda}\sqrt{\frac{\mu_0}{2}}+(\mbox{higher terms in}\; \mu_0)\label{lamdatilde}\\
\rho\sim\tilde{\rho}\sqrt{\frac{\mu_0}{2}}+(\mbox{higher terms in}\; \mu_0),\label{rhotilde}
\end{eqnarray} 
\end{subequations}
where the introduction of $\sqrt{2}$ is for later convenience and (\ref{rhotilde}) appears only for the type-$\Pcop$.
Here, the parameters are subject to
\begin{equation}
\tilde{\lambda}\tilde{\rho}\sin\psi=2D^2k',\label{Pcopconstraint}
\end{equation}
for the type-$\Pcop$ matching case, and 
\begin{equation}
\tilde{\lambda}^2=2D^2k',\label{Pcolconstraint}
\end{equation}
for the type-$\Pcol$ matching case.

Now,  we explicitly show the instanton ADHM data given by the large period limit of the calorons considered so far.
Apparently, the interval $I$ turns out to be infinitesimal at $\mu_0\to0$, so that the bulk Nahm data are given by the value at $s=0$.
This means that only the solution in the interval $\Ione$ is relevant.
Hence,  the instanton data obtained from the type-$\Pcop$ matching conditions with the non-standard monopole type bulk solution (\ref{non-m}) are
\begin{eqnarray}
\Delta=
\left(
\begin{array}{cc}
\tilde{\lambda} & \tilde{\rho}(\cos\psi+\sin\psi e_2)\\
D(-\sin\phi e_1+\cos \phi e_3) & Dk'(\cos\phi e_1+\sin \phi e_3)\\
Dk'(\cos\phi e_1+\sin \phi e_3) & D(\sin\phi e_1-\cos \phi e_3)\\
\end{array}
\right)\nonumber\\
+
\left(
\begin{array}{cc}
0 & 0\\
x+d & 0\\
0 & x+d \\
\end{array}
\right),\label{inst from Pcop-non-M}
\end{eqnarray}
while with the monopole type bulk solution (\ref{monopole}) are
\begin{equation}
\Delta=
\left(
\begin{array}{cc}
\tilde{\lambda} & \tilde{\rho}(\cos\psi+\sin\psi e_2)\\
D e_3 & Dk' e_1\\
Dk' e_1 & -D e_3\\
\end{array}
\right)+
\left(
\begin{array}{cc}
0 & 0\\
x+d & 0\\
0 & x+d \\
\end{array}
\right),\label{inst from Pcop-M}
\end{equation}
where $d=d^{(1)}_0+d^{(1)}_j e_j$ and the condition (\ref{Pcopconstraint}) holds.
For the latter case, we have performed the gauge transformation to be $\phi=0$.
Similarly, we find the instanton data from type-$\Pcol$ matching conditions are analogous to  (\ref{inst from Pcop-non-M}) and (\ref{inst from Pcop-M}),
but replaced the first rows with
\begin{equation}
W=\left(\tilde{\lambda}, \tilde{\lambda}e_2\right)
\end{equation}
along with (\ref{Pcolconstraint}).

Finally, we comment on the large period limit of type-O matching conditions.
From (\ref{matchingmatrixO}) and (\ref{O-components}), we find  $D_m=0$, which leads to a pure gauge configuration.
Hence, there is no instanton limit in this case.
If we need the instanton data from this type of matching, we have to take through the trivial holonomy calorons.

\section{Conclusion}

In this paper, we have constructed the analytic Nahm data of $\SUtwo$ calorons of instanton charge $k=2$. 
Although a part of analytic data has already been  considered in  \cite{nogradi, BNvB03, BNvB04, harland} along with numerical studies,
 this paper presents the Nahm data in its  appropriate form to investigate  the moduli space $\mathcal{M}_2$.
The exact bulk data are given in (\ref{Nahmansatz}) (\ref{monopole}),  (\ref{a and b}) and (\ref{non-m}).
In particular, we have found that the Nahm data of (non-M, non-M) with type-$\Pcop$ matching conditions gives
the case of maximal number of moduli parameters, 16, which is $\mathrm{dim}\mathcal{M}_2$.
In addition, the Nahm data of these forms are suitable for considering the large scale and the large period limit of the calorons.

It is expected that there exist higher charge ($k>2$) calorons with special symmetry, as symmetric monopoles.
In order to search for such calorons,  the non-standard monopole type solutions like (\ref{non-m}) to the Nahm equation will play a significant role,
as we have seen in the $k=2$ calorons.
It will be suggested that the comprehensive study on the analytic solutions to Nahm equations
provides further insight into the relationship between calorons, monopoles, and  instantons.

\section*{Acknowledgement}
The authors wish to thank Dr.~Derek Harland for helpful comments.
They are also grateful to Dr.~Benoit Charbonneau for informing us about the references.

\end{document}